\documentclass[reqno,12pt]{amsproc}

\begin{document}


\newtheorem{thm}{Theorem}[section]

\def\eqref#1{(\ref{#1})}
\def\eqrefs#1#2{(\ref{#1}) and (\ref{#2})}
\def\eqsref#1#2{(\ref{#1}) to (\ref{#2})}

\def\Eqref#1{Eq.~(\ref{#1})}
\def\Eqrefs#1#2{Eqs.~(\ref{#1}) and (\ref{#2})}
\def\Eqsref#1#2{Eqs.~(\ref{#1}) to (\ref{#2})}

\def\Ref#1{Ref.\cite{#1}}
\def\Refs#1{Refs.\cite{#1}}

\def\EQ #1\doneEQ{\begin{equation} #1 
\end{equation}}
\def\mEQ #1\donemEQ{\begin{multline} #1 
\end{multline}}
\def\sEQ #1\donesEQ{\begin{equation}\begin{split} #1 
\end{\split}\end{equation}}
\def\EQs #1\doneEQs{\begin{align} #1 
\end{align}}
\def\cEQs #1\donecEQs{\begin{gather} #1 
\end{gather}}

\def\eqtext#1{\hbox{\rm{#1}}}


\def\fewquad{\qquad\qquad}
\def\severalquad{\qquad\fewquad}
\def\manyquad{\qquad\severalquad}
\def\manymanyquad{\manyquad\manyquad}

\def\sub#1{
\setbox1=\hbox{{$\scriptscriptstyle #1$}} 
\dimen1=0.6\ht1
\mkern-2mu \lower\dimen1\box1 \hbox to\dimen1{\box1\hfill} }

\def\eqtext#1{\hbox{\rm{#1}}}

\def\mstrut{\mathstrut}
\def\hp#1{\hphantom{#1}}

\def\mixedindices#1#2{{\mstrut}^{\mstrut #1}_{\mstrut #2}}
\def\downindex#1{{\mstrut}^{\mstrut}_{\mstrut #1}}
\def\upindex#1{{\mstrut}_{\mstrut}^{\mstrut #1}}
\def\downupindices#1#2{{\mstrut}_{\mstrut #1}^{\hp{#1}\mstrut #2}}
\def\updownindices#1#2{{\mstrut}^{\mstrut #1}_{\hp{#1}\mstrut #2}}


\def\Parder#1#2{
\mathchoice{\partial{#1} \over\partial{#2}}{\partial{#1}/\partial{#2}}{}{} }
\def\parder#1{\partial/\partial{#1}}

\def\der#1{\partial\downindex{#1}}
\def\coder#1{\partial\upindex{#1}}
\def\D#1{D\downindex{#1}}
\def\coD#1{D\upindex{#1}}
\def\covder#1{\nabla\downindex{#1}}
\def\covcoder#1{\nabla\upindex{#1}}
\def\nder#1#2#3{\partial^{(#1)}\downupindices{#2}{#3}}

\def\cLie#1{{\mathcal L}_{#1}}
\def\cwLie#1#2{{\mathcal L}{}^{(#1)}_{#2}}
\def\sLie#1{\hat{\mathcal L}_{#1}}
\def\Lie#1{{\frak L}_{#1}}
\def\div{{\rm div\,}}


\def\g#1{g\downindex{#1}}
\def\flat#1{\eta\downindex{#1}}
\def\invflat#1{\eta\upindex{#1}}
\def\vol#1#2{\epsilon\downupindices{#1}{#2}}
\def\x#1#2{x\mixedindices{#1}{#2}}
\def\id#1#2{\delta\mixedindices{#2}{#1}}

\def\sodder#1#2{\sigma\downupindices{#1}{#2}}
\def\invsodder#1#2{\sigma\updownindices{#1}{#2}}
\def\gmatr#1#2{\gamma\mixedindices{#1}{#2}}

\def\Pgen#1#2{\mathcal{G}\downupindices{#1}{#2}}
\def\PLgen#1#2{\mathcal{S}\downupindices{#1}{#2}}


\def\curr#1#2#3{\Psi\mixedindices{\rm #1}{#3}}

\def\curl#1#2{\Theta\mixedindices{#1}{#2}}
\def\tcurl#1#2{\tilde\Theta\mixedindices{#1}{#2}}
\def\triv#1#2{\Upsilon\mixedindices{#1}{#2}}

\def\symm#1#2#3{Q\mixedindices{\rm #1}{#3}}
\def\symmgen{{\rm\bf X}_Q}

\def\adsymm#1#2{P\mixedindices{#1}{#2}}
\def\grad#1{\chi\upindex{#1}}

\def\H#1#2{\Psi\mixedindices{#1}{#2}}

\def\X#1#2#3#4{X\mixedindices{#1\hp{#2}#3}{\hp{#1}#2#4}}
\def\T#1#2{T\mixedindices{#1}{#2}}


\def\sphi#1#2{\phi\mixedindices{#1}{#2}}
\def\csphi#1#2{{\bar\phi}\mixedindices{#1}{#2}}
\def\sphider#1#2#3#4{\phi\mixedindices{#1\hp{#2}#3}{\hp{#1}#2#4}}
\def\csphider#1#2#3#4{{\bar\phi}\mixedindices{#1\hp{#2}#3}{\hp{#1}#2#4}}
\def\snphi#1#2#3#4{\phi\mixedindices{(#1)}{#2}\mixedindices{#3}{#4}}
\def\csnphi#1#2#3#4{\bar\phi\mixedindices{(#1)}{#2}\mixedindices{#3}{#4}}

\def\sder#1#2{\partial\mixedindices{#1}{#2}}
\def\sD#1#2{D\mixedindices{#1}{#2}}
\def\solD#1#2{\mathcal{D}\mixedindices{#1}{#2}}
\def\snder#1#2#3{\partial^{(#1)}\mixedindices{#2}{#3}}
\def\solnder#1#2#3{\mathcal{D}^{(#1)}\mixedindices{#2}{#3}}

\def\covsder#1#2{\nabla\mixedindices{#1}{#2}}

\def\sphiparder#1#2#3#4{\partial_\phi
\mixedindices{#1\hp{#2}#3}{\hp{#1}#2#4}}
\def\csphiparder#1#2#3#4{\partial_{\bar\phi}
\mixedindices{#1\hp{#2}#3}{\hp{#1}#2#4}}

\def\sphisol#1#2{\varphi\mixedindices{#1}{#2}}
\def\csphisol#1#2{{\bar\varphi}\mixedindices{#1}{#2}}
\def\sphidersol#1#2#3#4{\varphi\mixedindices{#1\hp{#2}#3}{\hp{#1}#2#4}}
\def\csphidersol#1#2#3#4{{\bar\varphi}\mixedindices{#1\hp{#2}#3}{\hp{#1}#2#4}}

\def\spinor#1#2#3{{#1}\mixedindices{#2}{#3}}
\def\cspinor#1#2#3{\bar{#1}\mixedindices{#2}{#3}}

\def\FE#1#2{\Delta\mixedindices{#1}{#2}}
\def\cFE#1#2{\bar\Delta\mixedindices{#1}{#2}}

\def\twos{{2s}}
\def\ind#1#2{{\mathbf #1}_{#2}}
\def\indsymm{\mathfrak{S}}

\def\jet#1{\phi^{[#1]}}
\def\xtrans#1{\zeta\upindex{#1}}
\def\sphitrans#1{\varrho\downindex{#1}}


\def\KS#1#2{\kappa\mixedindices{#1}{#2}}
\def\cKS#1#2{{\bar\kappa}\mixedindices{#1}{#2}}

\def\KV#1#2{\xi\mixedindices{#1}{#2}}
\def\othKV#1#2{\zeta\mixedindices{#1}{#2}}
\def\cKV#1#2{{\bar\xi}\mixedindices{#1}{#2}}
\def\cothKV#1#2{\bar\zeta\mixedindices{#1}{#2}}
\def\KY#1#2{Y\updownindices{#1}{#2}}
\def\othKY#1#2{\Upsilon\updownindices{#1}{#2}}
\def\cKY#1#2{\bar Y\updownindices{#1}{#2}}
\def\rKY#1#2{{\tilde Y}\updownindices{#1}{#2}}

\def\prodsKY#1#2#3{Y^+_{(#1)}\updownindices{#2}{#3}}
\def\prodKY#1#2#3{Y_{#1}\updownindices{#2}{#3}}
\def\prodKV#1#2#3{\varrho_{#1}\mixedindices{#2}{#3}}

\def\prodx#1#2#3{x^{(#1)}\mixedindices{#2}{#3}}

\def\w#1#2{\omega\downupindices{#2}{#1}}
\def\cw#1#2{\bar\omega\downupindices{#2}{#1}}


\def\so#1#2{o\mixedindices{#1}{#2}}
\def\si#1#2{\iota\mixedindices{#1}{#2}}
\def\cso#1#2{{\bar o}\mixedindices{#1}{#2}}
\def\csi#1#2{{\bar\iota}\mixedindices{#1}{#2}}
\def\l#1{\ell\upindex{#1}}
\def\n#1{n\upindex{#1}}
\def\m#1{m\upindex{#1}}
\def\cm#1{\bar m\upindex{#1}}

\def\a#1#2{\alpha\mixedindices{#1}{#2}}
\def\b#1#2{\beta\mixedindices{#2}{#1}}
\def\ca#1#2{\bar\alpha\mixedindices{#1}{#2}}
\def\cb#1#2{\bar\beta\mixedindices{#2}{#1}}
\def\ta#1{\tilde\alpha\upindex{#1}}
\def\tb#1{\tilde\beta\downindex{#1}}

\def\t#1#2{t\downupindices{#2}{#1}}
\def\r#1#2{r\mixedindices{#1}{#2}}
\def\u#1#2{u\downupindices{#2}{#1}}
\def\z#1#2{n\downupindices{#2}{#1}}

\def\fourspin#1#2{\begin{pmatrix} 
#1\\ 
#2 \end{pmatrix}}


\def\F#1#2{F\downupindices{#2}{#1}}
\def\duF#1#2{{*F}\downupindices{#2}{#1}}
\def\sF#1#2{F^+\downupindices{#2}{#1}}
\def\aF#1#2{F^-\downupindices{#2}{#1}}

\def\C#1#2{C\downupindices{#2}{#1}}
\def\duC#1#2{{*C}\downupindices{#2}{#1}}
\def\sC#1#2{C^+\downupindices{#2}{#1}}
\def\aC#1#2{C^-\downupindices{#2}{#1}}


\def\M#1#2{\psi\mixedindices{#1}{#2}}
\def\sM#1#2{\psi^+\mixedindices{#1}{#2}}
\def\aM#1#2{\psi^-\mixedindices{#1}{#2}}
\def\duM#1#2{{*\psi}\mixedindices{#1}{#2}}
\def\hM#1#2#3{\psi^{#1}\mixedindices{#2}{#3}}

\def\ad{{}^\dagger}


\def\c#1#2{{{#1}\choose{#2}}}
\def\quant#1#2{C^{#1}_{#2}}

\def\vs#1#2{{\mathcal V}\mixedindices{#2}{#1}}
\def\ks#1#2{{\mathcal K}\mixedindices{#2}{#1}}
\def\qvs#1#2{{\mathcal N}\mixedindices{#2}{#1}}

\def\i{{\rm i}}

\def\const{{\rm const}}

\def\cc{\ c.c.}

\def\sfrac#1#2{\tfrac{#1}{#2}}

\def\proj{\tfrac{1}{2}(\mathbf{1}\mp\i*)}

\def\Rnum{{\mathbb R}}
\def\Cnum{{\mathbb C}}

\def\Kvec/{Killing vector}
\def\Kspin/{Killing spinor}
\def\Kten/{Killing tensor}
\def\KYten/{Killing-Yano tensor}

\def\Pgroup/{Poincar\'e group}
\def\Pcov/{Poincar\'e covariant}
\def\PL/{Pauli-Lubansk\'i}

\def\ie/{i.e.}
\def\eg/{e.g.}

\title[Symmetries and currents of free fields]
{ Symmetries and currents of massless \\
neutrino fields, electromagnetic and graviton fields }


\author{Stephen Anco}
\address{Department of Mathematics, 
Brock University \\
St. Catharines, ON L2S 3A1 Canada}
\email{sanco@brocku.ca}

\author{Juha Pohjanpelto}
\address{Department of Mathematics, 
Oregon State University \\
Corvallis, OR 97331-4605 USA}
\email{juha@math.orst.edu}

\begin{abstract} 
A recent complete, explicit classification of all locally constructed 
symmetries and currents for free spinorial massless spin $s$ fields
on Minkowski space is summarized and extended to give 
a classification of all covariant symmetry operators and conserved tensors. 
The results, for physically interesting cases, are also presented 
in tensorial form for electromagnetic and graviton fields ($s=1,2$) 
and in Dirac 4-spinor form for neutrino fields ($s=\sfrac{1}{2}$). 
\end{abstract}

\maketitle

\section{ Introduction}

One of the earliest applications of 
group theory in the foundations of both classical and quantum field theory
was to the study of the fundamental linear spinorial equations
for free relativistic fields on Minkowski space
\cite{Penrosebook,Waldbook}. 
These field equations arise in a natural group theoretical manner
by providing unitary irreducible representations modulo a sign 
of the \Pgroup/ --- the isometry group of Minkowski space ---
realized on spinorial fields on spacetime. 
As shown by Wigner and Bargmann \cite{WignerBargmann}, 
the representations are characterized in terms of 
mass $m\ge 0$ and spin $s=0,\sfrac{1}{2},1,\sfrac{3}{2},2,\ldots$ 
of the field,
which are given by eigenvalues of the Casimir operators of 
the Lie algebra of the \Pgroup/. 
In particular, the square of the translation operator yields $m^2$
while the square of the (\PL/) spin operator yields $s(s+1)m^2$
for massive fields. 
The spin for massless fields has a special characterization
given by the magnitude of the helicity 
$\pm s=0,\pm\sfrac{1}{2},\pm 1,\pm\sfrac{3}{2},\pm 2,\ldots$ 
which arises from an equality between 
the translation operator and spin operator 
holding for irreducible representations when $m=0$. 

The most important cases of physical interest are
the spinorial fields with zero mass $m=0$ 
and nonzero spin $s=\sfrac{1}{2},1,2$,
respectively describing neutrino fields, 
electromagnetic fields and graviton fields (\ie/, linearized gravitation). 
Gravitino fields, described by $m=0$ and $s=\sfrac{3}{2}$,
are of theoretical interest in supersymmetric field theory. 
Due to their linear nature, 
all these fields have a rich structure of conserved currents and symmetries, 
which have interesting physical applications:
currents provide conserved quantities associated with 
the propagation of the fields on spacetime,
while symmetries lead to invariant solutions 
and are connected with separation of variables for the field equations. 

In recent work \cite{currentspaper,symmetrypaper}
by means of spinorial methods, 
we have obtained a complete, explicit classification of
all locally constructed spinorial symmetries and currents 
for massless fields of every spin $s\ge \sfrac{1}{2}$,
extending some earlier results \cite{maxwellsymmops,maxwellpaper}
obtained for the electromagnetic case $s=1$. 
As this classification uses the spinorial formulation of the field equations,
the symmetries and currents are derived in 
a gauge invariant and coordinate invariant spinor form. 
For physical applications, however, 
a tensorial form for integer spin fields 
and a Dirac 4-spinor form for half-integer spin fields
is the most appropriate formulation. 

In this paper we present the symmetries and currents 
in tensorial form for electromagnetic and graviton fields
and in Dirac 4-spinor form for neutrino and gravitino fields. 
In addition, we extend our previous results to give 
a complete classification of all \Pcov/ 
conserved tensors and symmetry operators 
for massless spinorial fields of every spin $s\ge \sfrac{1}{2}$.
Throughout we use the index notation and conventions of \Ref{Penrosebook}.

\section{ Spin s symmetries and currents }

On Minkowski space $M=(\Rnum^4,\flat{ab})$, 
recall that the Pauli spin matrices (and identity matrix) 
$\sodder{a}{AA'}$ provide an isomorphism between
the tangent space of $M$ 
and the space of real spinorial vectors over spinor space 
$(\Cnum^2,\vol{AB}{})$,
where $\flat{ab}$ is the Minkowski metric 
and $\vol{AB}{}$ is the spin metric, 
related by $\flat{ab}=\sodder{a}{AA'}\sodder{b}{BB'}\vol{AB}{}\vol{A'B'}{}$.
Hereafter we will omit $\sodder{a}{AA'}$ wherever convenient 
and simply write $a=AA'$ to identify vector and tensor fields
with vectorial and tensorial spinor fields on $M$. 

Massless spin $s$ fields are described by symmetric spinor fields 
$\sphi{}{A_1\cdots A_\twos}(x)$ on $M$ satisfying the field equation
\EQ
\FE{}{A'A_2\cdots A_\twos}\equiv 
\sder{A_1}{A'} \sphi{}{A_1\cdots A_\twos}(x) =0 , 
\label{spinseq}
\doneEQ
where $\sder{A}{A'}$ denotes the spinorial coordinate derivative operator
associated with standard Minkowski coordinates $\x{CC'}{}$. 
The vector space of $C^\infty_0$ solutions of \eqref{spinseq} defines
an irreducible representation of the double cover ISL$(2,\Cnum)$ 
of the \Pgroup/ of $M$,
with the group action generated by Lie derivatives 
with respect to \Kvec/s $\KV{c}{}$ on $M$, 
\EQ
\Lie{\KV{}{}} \sphi{}{A_1\cdots A_\twos}(x)
= \KV{CC'}{}\der{CC'}\sphi{}{A_1\cdots A_\twos}(x)
+ s\der{C'(A_1}\KV{C'C}{} \sphi{}{A_2\cdots A_\twos)C}(x) , 
\label{lieder}
\doneEQ
where $\Lie{\KV{}{}} \flat{ab}=0$. 
The Poincar\'e Lie algebra generated by $\Lie{\KV{}{}}$
comprises translations $\KV{a}{}\Pgen{a}{}$ 
and rotations/boosts $\coder{[a}\KV{b]}{}\Pgen{ab}{}$
defined by $\tfrac{1}{\i}\Lie{\KV{}{}}$ via the corresponding \Kvec/s
($\KV{a}{}=\const$, $\coder{[a}\KV{b]}{}=\const$, respectively). 
The \PL/ spin operator is defined by 
$\PLgen{a}{}=\vol{a}{bcd} \Pgen{b}{}\Pgen{cd}{}$. 
On $C^\infty_0$ solutions of the field equation, 
these are self-adjoint operators that satisfy 
$\Pgen{a}{}\Pgen{}{a}= \PLgen{a}{}\PLgen{}{a}= \PLgen{a}{}\Pgen{}{a}= 0$
and $\PLgen{a}{}=-s\Pgen{a}{}$, 
from which the helicity of $\sphi{}{A_1\cdots A_\twos}(x)$ 
is defined to be $-s$. 
A similar discussion applies to 
the complex conjugate massless spin $s$ field
$\csphi{}{A_1'\cdots A_\twos'}(x)$ satisfying
$\cFE{}{AA_2'\cdots A_\twos'}
= \sder{A_1'}{A} \csphi{}{A_1'\cdots A_\twos'}(x) =0$,
with helicity $+s$ as defined by 
the equality $\PLgen{a}{}=s\Pgen{a}{}$
holding on $C^\infty_0$ solutions of this field equation. 

The field equation \eqref{spinseq} possesses an important local solvability
property by which, for each $q\ge 1$, 
the values of $\sphi{}{A_1\cdots A_\twos}(x_o)$ 
and all of its symmetrized derivatives 
$\sder{C_1'}{(C_1}\cdots\sder{C_p'}{C_p}\sphi{}{A_1\cdots A_\twos)}(x_o)$ 
for $p\le q$ at any given point $\x{AA'}{o}$ in $M$
are freely specifiable data on solutions, 
as explained by Penrose \cite{Penrosebook}
using the notion of ``exact set of fields''. 
Thus, it is convenient to work with the associated coordinate space,
\EQ
J^q_\Delta(\phi) \equiv \{(
\x{CC'}{},\sphi{}{A_1\cdots A_{2s}},\sphider{}{(A_1\cdots A_{2s},}{C'_1}{C_1)},
\ldots, \sphider{}{(A_1\cdots A_{2s},}{C'_1\cdots C'_q}{C_1\cdots C_q)} )\} ,
\label{Jpoint}
\doneEQ
$0\le q\le \infty$,
known as the solution jet space of the field equation \eqref{spinseq},
where a point in $J^q_\Delta(\phi)$ corresponds to 
the values of the field and all symmetrized derivatives of the field
up to order $q$ at a point in $M$. 
This is a subspace of the full jet space 
$J^q(\phi) \supset J^q_\Delta(\phi)$ 
whose coordinates are defined by 
$\x{CC'}{}$, $\sphi{}{A_1\cdots A_{2s}}$,
$\sphider{}{A_1\cdots A_{2s},}{C'_1\cdots C'_p}{C_1\cdots C_p}$, 
$1\le p\le q$. 
In the sequel we will employ a multi-index notation
and write
\EQ
\sphider{}{\ind{A}{2s},}{\ind{C}{p}'}{\ind{C}{p}} 
= \sphider{}{A_1\cdots A_{2s},}{C_1'\cdots C_p'}{C_1\cdots C_p} ,\quad
\sphi{\ind{C}{p}'}{\ind{C}{2s+p}} 
= \sphider{}{(\ind{C}{2s},}{\ind{C}{p}'}{\ind{C}{p,2s})} ,\quad
p\ge 0, 
\doneEQ
with multi-indices defined to be completely symmetric 
in their constituent indices:
$\ind{B}{p}=(B_1\cdots B_p)$, $\ind{B}{p,q}=(B_{1+q}\cdots B_{p+q})$. 
We will use the convention that a multi-index with $p=0$
stands for an empty set containing no index. 

We let $\sD{}{CC'}$ denote the total derivative operator 
with respect to $\x{CC'}{}$ on $J^\infty(\phi)$ 
and write $\solD{}{CC'}$ for its restriction to $J^\infty_\Delta(\phi)$ 
given by
\EQ
\solD{C'}{C} = 
\sder{C'}{C} +\sum\nolimits_{q\geq 0}
( \sphi{\ind{A}{q}'C'}{\ind{A}{2s+q}C}
\sphiparder{}{}{\ind{A}{2s+q}}{\ind{A}{q}'}
+\cc ) ,
\doneEQ
where
$\sphiparder{}{}{\ind{A}{2s+p}}{\ind{A}{p}'}$
is the partial derivative operator with respect to 
$\sphi{\ind{A}{p}'}{\ind{A}{2s+p}}$
and $\cc$
denotes the complex conjugate of the preceeding term. 
We write higher order symmetrized derivatives as
$\snder{p}{\ind{C}{p}'}{\ind{C}{p}}
= \sder{(C_1'}{(C_1}\cdots\sder{C_p')}{C_p)}$
and $\solnder{p}{\ind{C}{p}'}{\ind{C}{p}}
= \solD{(C_1'}{(C_1}\cdots\solD{C_p')}{C_p)}$. 
Note that we can lift the Lie derivative \eqref{lieder}
for any \Kvec/ $\KV{CC'}{}$ 
to define an operator $\Lie{\KV{}{}}$ on $J^\infty_\Delta(\phi)$ 
given by 
$\Lie{\KV{}{}} \sphi{\ind{A}{p}'}{\ind{A}{2s+p}}
=
-\KV{C'}{C}\sphi{\ind{A}{p}'C'}{\ind{A}{2s+p}C}
+ (s+\tfrac{p}{2})\KV{\hp{(A_{2s+p}}C}{(A_{2s+p}} 
\sphi{\ind{A}{p}'}{\ind{A}{2s+p-1})C} 
- \tfrac{p}{2}\cKV{\hp{C'}(A_{p}'}{C'}
\csphi{\ind{A}{p-1}')C'}{\ind{A}{2s+p}}$
where $\KV{}{BC}=\der{C'B}\KV{C'}{C}$.

\subsection{ Conformal Killing vectors and Killing-Yano tensors }

The classification of local symmetries and local currents of 
massless spin $s$ fields 
given in \Refs{currentspaper,symmetrypaper}
relies on the properties of \Kspin/s, 
which are spinorial generalizations of \Kvec/s related to twistors
\cite{Penrosebook}.
For the results presented here, 
we need \Kspin/s of two types. 
A real spinor function $\KV{A'}{A}(x)$ 
satisfying 
$\sder{(B'}{(B} \KV{A')}{A)} =0$
represents a conformal \Kvec/ $\KV{a}{}$ 
\cite{Waldbook,Penrosebook}, 
which generates a conformal isometry of Minkowski space. 
A symmetric spinor function $\KY{A'B'}{}(x)$
satisfying
$\sder{(C'}{C} \KY{A'B')}{}=0$
represents a conformal \KYten/ $\KY{ab}{} = \vol{}{AB}\KY{A'B'}{}$ 
\cite{DietzRudiger,Penrosebook} 
that is self-dual,
$*\KY{ab}{}=\i\KY{ab}{}$,
where $*$ denotes the Hodge dual operator.
These \Kspin/s have a direct generalization 
$\othKV{\ind{A}{k}'}{\ind{A}{k}}(x)$ 
and $\othKY{\ind{A}{2k}'}{}(x)$
to ones of any rank $k\ge 1$. 
Their explicit form is given by polynomials in $\x{CC'}{}$ 
of degree up to $2k$, 
\EQs
&
\othKV{\ind{A}{k}'}{\ind{A}{k}}
= \sum_{0\le p\le q\le k} 
\spinor{\alpha}{\ind{B}{p}'(\ind{A}{k-q,q}'}{\ind{B}{q}(\ind{A}{k-p,p}}
\prodx{q}{\ind{A}{q}')\ind{B}{q}}{} \prodx{p}{}{\ind{A}{p})\ind{B}{p}'} 
+ \cc ,
\label{CKtens}\\
&
\othKY{\ind{A}{2k}'}{} 
= \sum_{0\le p\le 2k} 
\spinor{\beta}{(\ind{A}{2k-p,p}'}{\ind{B}{p}}
\prodx{p}{\ind{A}{p}')\ind{B}{p}}{} ,
\label{CKYtens}
\doneEQs
where 
$\prodx{p}{\ind{C}{p}'}{\ind{C}{p}}= \x{(C_1'}{C_1}\cdots\x{C_p')}{C_p}$,
with the coefficients 
$\spinor{\alpha}{\ind{B}{p}'\ind{A}{k-q}'}{\ind{B}{q}\ind{A}{k-p}}$
and $\spinor{\beta}{\ind{A}{2k-p}'}{\ind{B}{p}}$
being arbitrary constant spinors. 
There are respectively
\EQ
(k+1)^2 (k+2)^2 (2k+3)/12 ,\quad
(2k+1) (2k+2) (2k+3)/3
\doneEQ
linearly independent \Kspin/s \eqrefs{CKtens}{CKYtens} over the reals. 

An important property of these \Kspin/s is that 
they possess a factorization into sums of symmetrized products of 
conformal \Kvec/s $\KV{A'}{A}$ 
and conformal \KYten/s $\KY{A'B'}{}$:
\EQ
\othKV{\ind{A}{k}'}{\ind{A}{k}}
= \sum\nolimits_\KV{}{} \KV{A'_1}{(A_1} \cdots \KV{A'_k}{A_k)} ,\quad
\othKY{\ind{A}{2k}'}{} 
= \sum\nolimits_\KY{}{} \KY{(A'_1A'_2}{} \cdots \KY{A'_{2k-1}A'_{2k})}{} .
\label{KYproduct}
\doneEQ
This is a consequence of the more general factorization 
of \Kspin/s into sums of symmetrized products of twistors and dual-twistors,
holding in Minkowski space.

\subsection{ Symmetries }

From a group theoretical perspective, 
a local symmetry of the massless spin $s$ field equation \eqref{spinseq}
is a one-parameter ($\varepsilon$) local transformation group \cite{book}
on the coordinate space $J^\infty(\phi)$ 
that preserves the contact ideal
(\ie/, derivative relations among the coordinates \cite{book,Olverbook})
and maps solutions $\sphi{}{\ind{A}{2s}}(x)$ into solutions. 
It is well known that the infinitesimal action of any such transformation
on $\sphi{}{\ind{A}{2s}}(x)$ is the same as one in which 
there is no motion on $\x{CC'}{}$, 
\EQ
\x{CC'}{} \rightarrow \x{CC'}{} ,\quad
\sphi{}{\ind{A}{2s}} \rightarrow 
\sphi{}{\ind{A}{2s}} +\varepsilon \symm{}{}{\ind{A}{2s}}(x,\jet{r}) 
+O(\varepsilon^2) ,
\label{trans}
\doneEQ
with the prolongation 
$\sphi{\ind{A}{p}'}{\ind{A}{2s+p}} \rightarrow
\sphi{\ind{A}{p}'}{\ind{A}{2s+p}} 
+\varepsilon \solnder{p}{\ind{A}{p}'}{(\ind{A}{p}}
\symm{}{}{\ind{A}{2s,p})}(x,\jet{r}) 
+O(\varepsilon^2)$ 
for $p\ge 1$. 
The spinor function $\symm{}{}{\ind{A}{2s}}(x,\jet{r})$ is called
the symmetry characteristic of the local transformation group \eqref{trans}
and satisfies the determining equation
\EQ
\solD{A_\twos}{A'} \symm{}{}{\ind{A}{2s}}(x,\jet{r}) =0 .
\label{symmeq}
\doneEQ
Here $\jet{r}$ denotes the set of coordinates 
$\sphi{\ind{A}{p}'}{\ind{A}{2s+p}},\csphi{\ind{A}{p}}{\ind{A}{2s+p}'}$, 
with $0\le p\le r$. 
The infinitesimal generator of the resulting local transformation group
(defined by formal exponentiation \cite{book} of the generator)
is given by 
\EQ
\symmgen = \symm{}{}{\ind{A}{2s}} \sphiparder{\ind{A}{2s}}{}{}{} , 
\label{symm}
\doneEQ
which we will call a local spin $s$ symmetry of order $r$. 
More geometrically, 
a local symmetry can be understood \cite{Olverbook} 
to be a tangent vector field 
on the solution jet space 
$J^\infty_\Delta(\phi) \subset J^\infty(\phi)$
that preserves the contact ideal
associated with the coordinates. 

If $\symm{}{}{\ind{A}{2s}}$ depends only on $\x{CC'}{}$, 
so that $\symm{}{}{\ind{A}{2s}}(x)$ is 
a solution of \eqref{spinseq}, 
we call $\symmgen$ an elementary spin $s$ symmetry. 
A spin $s$ symmetry $\symmgen$ is a classical point symmetry 
\cite{book,Olverbook}
if it has the form 
$\symm{}{}{\ind{A}{2s}}(x,\jet{1})= 
\xtrans{CC'}\sphi{}{\ind{A}{2s}CC'} + \sphitrans{\ind{A}{2s}}$
for some spinor functions 
$\xtrans{CC'}(x,\jet{0}),\sphitrans{\ind{A}{2s}}(x,\jet{0})$. 
Allowing for complexification, 
the point symmetries admitted by 
massless spin $s$ fields
consist of the scaling and duality rotation symmetries
\EQ
\symm{S}{}{\ind{A}{2s}}(\jet{0}) = \sphi{}{\ind{A}{2s}} ,\quad
\symm{S}{}{\ind{A}{2s}}(\i\jet{0}) = \i\sphi{}{\ind{A}{2s}} , 
\label{scalingsymm}
\doneEQ
as well as the spacetime symmetries 
\EQ
\symm{K}{}{\ind{A}{2s}}(\KV{}{},\jet{1}) 
= \sLie{\KV{}{}}\sphi{}{\ind{A}{2s}} ,\quad
\symm{K}{}{\ind{A}{2s}}(\KV{}{},\i\jet{1}) 
= \i\sLie{\KV{}{}}\sphi{}{\ind{A}{2s}} ,
\label{spacetimesymm}
\doneEQ
arising from the action of
the group of conformal isometries of Minkowski space 
generated by conformal \Kvec/s $\KV{c}{}$,
where the operator 
\EQ
\sLie{\KV{}{}} = \Lie{\KV{}{}} + \tfrac{1-s}{4} \div\KV{}{} ,\quad
\div\KV{}{}\equiv \der{a}\KV{a}{}
\doneEQ
is, geometrically, a conformally-weighted Lie derivative
\cite{Penrosebook,currentspaper}.

Massless spin $s$ fields, remarkably, 
also admit non-classical local symmetries involving conformal \KYten/s,
given by 
\EQ
\symm{C}{}{\ind{A}{2s}}(\KY{}{},\jet{2s}) 
= \sum_{0\le p\le 2s} \tfrac{4s-p+1}{4s+1}\tbinom{2s}{p}
\snder{p}{}{\ind{B}{p}'(\ind{A}{p}} \KY{\ind{B}{4s}'}{}
\csphi{}{|\ind{B}{4s-p,p}'|\ind{A}{2s-p,p})} ,
\label{chiralsymm}
\doneEQ
where $\KY{\ind{B}{4s}'}{}$ is any self-dual conformal \KYten/ of rank $2s$.
Symmetries of this type were first found in tensorial form 
in the electromagnetic case $s=1$ by Fushchich and Nikitin
\cite{newsymm,FushchichNikitin,survey}.
The generalization \eqref{chiralsymm} for all $s\ge \tfrac{1}{2}$
was derived in \Ref{symmetrypaper}.
We call \eqref{chiralsymm} 
chiral symmetries of order $2s$
since $\symm{C}{}{\ind{A}{2s}}$ depends on 
the positive helicity spin $s$ field $\csphi{}{\ind{A}{2s}'}$,
in contrast to the dependence of the spacetime symmetries 
$\symm{K}{}{\ind{A}{2s}}$ 
on the opposite helicity spin $s$ field $\sphi{}{\ind{A}{2s}}$. 

We now state the main classification result for local spin $s$ symmetries. 
First, note that 
given any local spin $s$ symmetry $\symmgen$ of order $r\ge 0$,
we can obtain higher order symmetries by replacing $\jet{r}$
in $\symm{}{}{\ind{A}{2s}}(x,\jet{r})$
with repeated Lie derivatives $(\sLie{\KV{}{}})^n\jet{r}$
for any conformal \Kvec/ $\KV{c}{}$,
since $\sLie{\KV{}{}}\sphi{}{\ind{A}{2s}}(x)$ is a solution of 
the massless spin $s$ field equation 
whenever $\sphi{}{\ind{A}{2s}}(x)$ is one. 
We denote by 
$\symm{}{}{\ind{A}{2s}}(\KV{(n)}{};x,\jet{r+n})$
the resulting symmetry characteristic 
for $n\ge 0$.

\begin{thm}\label{symmetryclass}
Every local symmetry \eqref{symm}
of the massless spin $s$ field equation \eqref{spinseq}
is a sum of an elementary symmetry and a linear symmetry that is given by,
to within a scaling and duality rotation, 
a sum of spacetime symmetries \eqref{spacetimesymm},
chiral symmetries \eqref{chiralsymm},
and their higher order extensions
$$
\sum_{n\ge 0,\ \KV{}{},\KY{}{}}
\symm{K}{}{\ind{A}{2s}}(\KV{(n)}{};\KV{}{},\jet{1+n})
+ \i\symm{K}{}{\ind{A}{2s}}(\KV{(n)}{};\KV{}{},\jet{1+n})
+ \symm{C}{}{\ind{A}{2s}}(\KV{(n)}{};\KY{}{},\jet{2s+n}) 
$$
involving real conformal \Kvec/s $\KV{}{}$
and self-dual conformal \KYten/s $\KY{}{}$ of rank $2s$. 
\end{thm}

\subsection{ Currents }

A local conserved current of the massless spin $s$ field equation 
\eqref{spinseq} 
is real vector function on the coordinate space $J^\infty(\phi)$ 
that it is divergence free 
on all solutions $\sphi{}{\ind{A}{2s}}(x)$ of \eqref{spinseq}. 
Without loss of generality, it is convenient to restrict local currents
to be divergence-free vector functions 
$\curr{}{}{a}(x,\jet{r})$
on the solution jet space $J^\infty_\Delta(\phi)$, 
\EQ
\solD{a}{}\curr{}{}{a}(x,\jet{r}) =0 .
\label{curr}
\doneEQ
Consider a spacelike hyperplane $\Sigma$,
with a future timelike normal $\t{a}{}$. 
For any current $\curr{}{}{a}(x,\jet{r})$,
the associated conserved quantity 
for $C^\infty_0$ solutions $\sphi{}{\ind{A}{2s}}(x)$
is then 
$\int_\Sigma \t{a}{} \curr{}{}{a}(x,\jet{r}(x)) d^3x$
where $\t{a}{} \curr{}{}{a}(x,\jet{r}(x))$ 
is the conserved density expression. 
This quantity is finite and time-independent. 
Thus, a local current $\curr{}{}{a}(x,\jet{r})$ is considered trivial 
if it agrees with a curl 
$\curr{}{}{AA'}= \sD{B'}{A}\curl{}{A'B'} +\cc$
on $J^\infty_\Delta(\phi)$, 
for some symmetric spinor function $\curl{}{A'B'}(x,\jet{r})$,
since the resulting conserved quantity vanishes by Stokes' theorem. 
Consequently, two local currents are considered equivalent 
if their difference is a trivial current. 

The massless spin $s$ field equation \eqref{spinseq}
does not possess a local Lagrangian formulation
in terms of $\sphi{}{\ind{A}{2s}}$ and its derivatives
(and their complex conjugates). 
As a result, 
local spin $s$ currents do not arise from local spin $s$ symmetries
via Noether's theorem 
but instead are related to adjoint symmetries of 
the field equation \eqref{spinseq} as follows. 
A spin $s$ adjoint symmetry of order $r$ is a spinor function 
$\adsymm{\ind{A}{2s-1}'}{A}(x,\jet{r})$ 
that satisfies the adjoint of the symmetry determining equation \eqref{symmeq}
\EQ
\solD{A(A_\twos'}{}\adsymm{\ind{A}{2s-1}')}{A}(x,\jet{r}) =0 .
\label{adsymmeq}
\doneEQ
Every spin $s$ adjoint symmetry generates a local conserved current
through a homotopy integral formula 
\EQ
\H{}{AA'}(\adsymm{}{}) =
\int_0^1 d\lambda \csphi{}{A'\ind{A}{2s-1}'}
\adsymm{\ind{A}{2s-1}'}{A}(x,\lambda\jet{r}) 
+\cc
\label{curreq}
\doneEQ
which is derived from the adjoint relation 
between equations \eqrefs{adsymmeq}{symmeq}. 
Conversely, as shown in \Ref{currentspaper},
every local spin $s$ current \eqref{curr}
is equivalent to one given by the integral formula \eqref{curreq}
for some spin $s$ adjoint symmetry. 
Note when $\adsymm{\ind{A}{2s-1}'}{A}$ depends only on $\x{CC'}{}$,
so that $\adsymm{\ind{A}{2s-1}'}{A}(x)$ 
is a solution of the adjoint spin $s$ field equation,
we obtain the elementary, linear currents of 
the massless spin $s$ field equation \eqref{spinseq}. 

Quadratic currents depending on \Kvec/s
have long been known in the electromagnetic case $s=1$,
corresponding to conservation of 
energy, momentum, angular and boost momentum
given via the electromagnetic stress-energy tensor. 
Moreover, so-called zilch quantities 
for electromagnetic fields 
are known to arise in a similar fashion 
from Lipkin's zilch tensor \cite{survey}.
Analogous local currents and tensors are also known
in the graviton case $s=2$ \cite{zilch}.
Generalizations of these currents 
in spinorial form for all $s\ge \tfrac{1}{2}$ 
were first obtained in \Ref{currentspaper}, 
given by 
\cEQs
\curr{K}{}{AA'}(\othKV{}{},\jet{0}) = 
\othKV{\ind{A}{2s-1}\ind{A}{2s-1}'}{}
\csphi{}{A'\ind{A}{2s-1}'} \sphi{}{A\ind{A}{2s-1}} ,
\label{spacetimecurr} \\
\curr{Z}{}{AA'}(\KV{}{},\othKV{}{},\jet{1}) = 
\i\othKV{\ind{A}{2s-1}\ind{A}{2s-1}'}{}
\csphi{}{A'\ind{A}{2s-1}'} \sLie{\KV{}{}}\sphi{}{A\ind{A}{2s-1}} 
+\cc , 
\label{zilchcurr}
\donecEQs
for any real conformal \Kvec/s $\KV{CC'}{}$
and real conformal \Kten/s $\othKV{\ind{A}{2s-1}\ind{A}{2s-1}'}{}$ 
of rank $2s-1$. 
We will refer to \eqrefs{spacetimecurr}{zilchcurr}
as the spacetme currents and zilch currents.
These currents possess even parity under duality rotations 
of the spin $s$ field. 
Remarkably, 
the massless spin $s$ field equation also admits odd parity currents, 
first found in tensorial form in the electromagnetic case $s=1$ 
by Fushchich and Nikitin \cite{FushchichNikitin}
using non-invariant coordinate methods. 
These currents were generalized in \Ref{currentspaper} 
to all $s\ge \tfrac{1}{2}$ in spinorial form, 
\mEQ
\curr{C}{}{AA'}(\KV{}{},\KY{}{},\jet{1}) = 
( \KY{\ind{A}{2s}'\ind{B}{2s}'}{} \csphi{}{\ind{B}{2s}'A_\twos'A} +
\\
\tfrac{2s+1}{4s+1}\der{AA_\twos'}\KY{\ind{A}{2s}'\ind{B}{2s}'}{} 
\csphi{}{\ind{B}{2s}'} )
\sLie{\KV{}{}}\csphi{}{A'\ind{A}{2s-1}'}
+\cc
\label{chiralcurr}
\donemEQ
for any conformal \KYten/s $\KY{\ind{A}{4s}'}{}$ of rank $2s$
and any conformal \Kvec/s $\KV{CC'}{}$. 
Since \eqref{chiralcurr} is of opposite parity 
to \eqrefs{spacetimecurr}{zilchcurr}, 
we call \eqref{chiralcurr} the chiral currents. 

A complete classification of local spin $s$ currents arises
from $\H{}{AA'}(\adsymm{}{})$ 
by a classification of local spin $s$ adjoint symmetries
similarly to theorem~\ref{symmetryclass}. 
As was the case for local symmetries, 
given any local spin $s$ current $\curr{}{}{AA'}(x,\jet{r})$ of order $r\ge 0$,
we can replace $\jet{r}$ by repeated Lie derivatives 
$(\sLie{\KV{}{}})^n\jet{r}$ 
to obtain higher order currents,
which we will denote by $\curr{}{}{AA'}(\KV{(n)}{};x,\jet{r+n})$, $n\ge 0$. 

\begin{thm}\label{currentclass}
Every local current \eqref{curr} 
of the massless spin $s$ field equation \eqref{spinseq}
is equivalent to a sum of an elementary linear current 
and a quadratic current given by
a sum of spacetime currents \eqref{spacetimecurr},
zilch currents \eqref{zilchcurr},
chiral currents \eqref{chiralcurr},
and their higher order extensions
$$
\sum\limits_{n\ge 0,\ \KV{}{},\othKV{}{},\KY{}{}}
\curr{K}{}{a}(\KV{(n)}{};\othKV{}{},\jet{n})
+ \curr{Z}{}{a}(\KV{(n)}{};\KV{}{},\othKV{}{},\jet{1+n})
+ \curr{C}{}{a}(\KV{(n)}{};\KV{}{},\KY{}{},\jet{1+n})
$$
involving real conformal \Kvec/s $\KV{}{}$
and \Kten/s $\othKV{}{}$ of rank $2s-1$, 
and self-dual conformal \KYten/s $\KY{}{}$ of rank $2s$. 
\end{thm}

\subsection{ Covariant conserved tensors and symmetry operators }

We now extend the previous classification results to
covariant conserved tensors and symmetry operators 
of the massless spin $s$ field equation \eqref{spinseq}. 
To begin, recall a spinor function is said to be \Pcov/
if it transforms equivariantly under  
the double cover ISL$(2,\Cnum)$ of the \Pgroup/ acting on $\jet{r}$
and hence depends purely on the coordinates $\jet{r}$ 
and spin metric $\vol{AB}{}$. 
On the solution jet space $J^\infty_\Delta(\phi)$, 
a covariant conserved tensor 
$\T{A'\ind{B}{q}'}{A\ind{B}{p}}(\jet{r})$ of order $r$
is then a spinor function that is \Pcov/ 
and divergence free, 
$\solD{A}{A'} \T{A'\ind{B}{q}'}{A\ind{B}{p}}(\jet{r})=0$,
and a covariant symmetry operator 
$\X{}{\ind{A}{2s}}{\ind{B}{q}'}{\ind{B}{p}}(\jet{r})
\sphiparder{\ind{A}{2s}}{}{}{}$ of order $r$ 
is characterized by a spinor function that is \Pcov/
and satisfies the symmetry equation
$\solD{A_\twos}{A'} \X{}{\ind{A}{2s}}{\ind{B}{q}'}{\ind{B}{p}}(\jet{r}) =0$. 

By contracting any covariant conserved tensor or symmetry operator
with products of an arbitrary constant spinor $\KS{B}{}$ 
and its conjugate $\cKS{B'}{}$,
we obtain a local current or symmetry, respectively.
Conversely, 
if the \Kspin/s 
$\KV{C'}{C}$, $\othKV{\ind{A}{2s-1}'}{\ind{A}{2s-1}}$, $\KY{\ind{A}{4s}'}{}$
in any local current or symmetry are 
set to equal products of $\KS{B}{},\cKS{B'}{}$ and factored out, 
then we clearly obtain a covariant conserved tensor or symmetry operator. 
The classification theorems \ref{symmetryclass} and \ref{currentclass}
now lead (as shown with the methods of \Refs{currentspaper,symmetrypaper})
to the following results. 

\begin{thm}
Every covariant spin $s$ symmetry operator is 
a complex linear combination of 
spacetime and chiral symmetry operators, 
$$
\X{}{\ind{A}{2s}}{\ind{B}{p}'}{\ind{B}{p}}
= \sphi{\ind{B}{p}'}{\ind{B}{p}\ind{A}{2s}} ,\quad
\X{}{\ind{A}{2s}}{\ind{B}{4s+p}'}{\ind{B}{p}}
= \csphi{\ind{B}{4s+p}'}{\ind{B}{p}\ind{A}{2s}} ,\quad
\text{for $p\ge 0$,}
$$
in addition to the elementary operator
$\X{}{\ind{A}{2s}}{\ind{B}{2s}}{} = \id{\ind{A}{2s}}{\ind{B}{2s}}$. 
Every covariant spin $s$ conserved tensor is equivalent to 
a complex linear combination of 
the elementary tensor
$\T{A'B'}{A\ind{B}{2s-1}}= \vol{}{A'B'}\sphi{}{A\ind{B}{2s-1}}$,
and spacetime tensors, zilch tensors, and chiral tensors, 
\cEQs
\T{A'\ind{B}{2s+2p-1}'}{A\ind{B}{2s+2p-1}}
= \csphi{A'(\ind{B}{2s+p-1}'}{(\ind{B}{p,2s+p-1}}
\sphi{\ind{B}{p,2s+p-1}')}{\ind{B}{2s+p-1})A} ,\
\T{A'\ind{B}{2s+2p}'}{A\ind{B}{2s+2p}}
= \i\csphi{A'(\ind{B}{2s+p}'}{(\ind{B}{p+1,2s+p-1}}
\sphi{\ind{B}{p,2s+p}')}{\ind{B}{2s+p-1})A} ,
\notag\\
\T{A'\ind{B}{4s+2p+1}'}{A\ind{B}{2p+1}}
= \csphi{(\ind{B}{4s+p+1}'}{A(\ind{B}{p}}
\csphi{\ind{B}{p,4s+p+1}')A'}{\ind{B}{p+1,p})} ,\quad
\text{for $p\ge 0$,}
\notag
\donecEQs
in addition to their complex conjugates.
\end{thm}

\section{ Results for spin $s=1/2$, $1$, $3/2$, $2$ }

Real conformal \Kvec/s $\KV{a}{}=\KV{AA'}{}$ 
and self-dual conformal \KYten/s $\KY{ab}{}=\vol{}{AB}\KY{A'B'}{}$
satisfy the tensorial equations 
\EQ
\coder{(a} \KV{b)}{} = 
\tfrac{1}{4} \invflat{ab} \der{c} \KV{c}{} , \qquad
\coder{(a}\KY{b)d}{} = 
\tfrac{1}{3}\invflat{ab} \der{c}\KY{cd}{}
+\tfrac{1}{3}\invflat{d(a} \der{c}\KY{b)c}{} 
\label{KVeqKYeq}
\doneEQ
whose solutions are quadratic polynomials in the coordinates $\x{a}{}$, 
\EQs
& \KV{a}{} = 
\alpha_1\upindex{a} 
+\alpha_2\updownindices{ab}{} \x{}{b}
+\alpha_3 \x{a}{}
+\alpha_4\upindex{c} \x{}{c}\x{a}{}
- \tfrac{1}{2}\alpha_4\upindex{a} \x{c}{}\x{}{c},
\label{Kvecs}
\\
& \KY{ab}{} =
\beta_1\upindex{ab} 
+\beta_2\upindex{[a} \x{b]^+}{}
+\beta_3\upindex{c[a} \x{b]^+}{}\x{}{c} 
\label{KYtens}
\doneEQs
with constant coefficients (respectively, real and complex valued)
\EQ\label{consts}
\alpha_1\upindex{a}, 
\alpha_2\upindex{ab}= \alpha_2\upindex{[ab]}, 
\alpha_3, 
\alpha_4\upindex{c}, 
\beta_1\upindex{ab}= \beta_1\upindex{[ab]^+}, 
\beta_2\upindex{a}, 
\beta_3\upindex{ab}= \beta_3\upindex{[ab]^-},
\doneEQ
where we use $+/-$ superscripts to denote
self-/antiself- dual projections as defined by $\proj$. 
There are 15 linearly independent conformal \Kvec/s \eqref{Kvecs}
and 20 linearly independent self-dual conformal \KYten/s \eqref{KYtens} 
over the reals. 
Hereafter, we write 
$\cwLie{w}{\KV{}{}} = \cLie{\KV{}{}} -\tfrac{w}{4}\div\KV{}{}$
where $\cLie{\KV{}{}}=\sLie{\KV{}{}} -\tfrac{1}{4}\div\KV{}{}$ 
is the ordinary Lie derivative operator \cite{Penrosebook}
satisfying the Killing equation
$\cLie{\KV{}{}}\flat{ab} =\tfrac{1}{2}\flat{ab}\div\KV{}{}$.

\subsection{ Electromagnetic fields }
\label{spinone}

In tensorial form 
a spin $s=1$ field 
is represented by the electromagnetic field tensor
\EQ
\F{}{ab} = \vol{AB}{}\csphi{}{A'B'} + \cc 
\doneEQ
which is real, antisymmetric, 
and satisfies the Maxwell field equations
\EQ
\coder{a} \F{}{ab}(x) =\coder{a} \duF{}{ab}(x) =0 , 
\label{ME}
\doneEQ
where $*$ is the Hodge dual,
$\duF{}{ab}= \i\vol{AB}{}\csphi{}{A'B'} + \cc$. 
It is convenient to decompose $\F{}{ab}$ into 
its self-dual and antiself-dual parts 
\EQs
&
\sF{}{ab} =\tfrac{1}{2}(\F{}{ab}-\i\duF{}{ab}) 
= \vol{AB}{}\csphi{}{A'B'} ,
\\&
\aF{}{ab} =\tfrac{1}{2}(\F{}{ab}+\i\duF{}{ab}) =\overline{\sF{}{ab}}
= \vol{A'B'}{}\sphi{}{AB} .
\doneEQs

The electromagnetic scaling and duality rotation symmetries 
are given by 
$\symm{S}{}{ab} =\F{}{ab}$, 
$\symm{}{}{ab} =\duF{}{ab} =*\symm{S}{}{ab}$,
while the spacetime symmetries 
depending on real conformal \Kvec/s $\KV{c}{}$ 
have the form
\EQ
\symm{K}{}{ab} =
\cLie{\KV{}{}} \F{}{ab} ,\quad
\symm{}{}{ab} =
\cLie{\KV{}{}} \duF{}{ab} = *\symm{K}{}{ab} ,
\doneEQ
reflecting the invariance \cite{Waldbook,Penrosebook} of \eqref{ME}
under conformal scalings of $\flat{ab}$. 
The chiral symmetries are given by 
\mEQ
\symm{C}{}{ab} = \sum\nolimits_{\KY{}{}} (
\prodsKY{2}{de}{c[b} \der{a]} \coder{c} \sF{}{de}
+\tfrac{8}{5} \der{[a|}\prodsKY{2}{de}{c|b]^+} \coder{c}\sF{}{de} 
\\
+\tfrac{1}{5} \der{[a|}\coder{c} \prodsKY{2'}{de}{c|b]} \sF{}{de} )
+\cc
\donemEQ
which depend on self-dual conformal \KYten/s $\KY{ab}{}$, 
where we have introduced the product tensors
\EQ
\prodsKY{2}{cdef}{} = \KY{cd}{}\KY{ef}{} ,\quad
\prodsKY{2'}{cdef}{} = \KY{cd}{}\KY{ef}{} -4\KY{c[e}{}\KY{f]^+d}{}
\doneEQ
associated with terms arising in the factorization \eqref{KYproduct} 
of rank-two self-dual conformal \KYten/s
in tensorial form. 

The spacetime currents and zilch currents 
are given by 
\EQ
\curr{K}{}{a}= 
\KV{}{b} \sF{}{ac} \aF{bc}{} +\cc ,\quad
\curr{Z}{}{a}= \sum\nolimits_{\KV{}{}}
\i \KV{}{b} \aF{bc}{} \cLie{\KV{}{}}\sF{}{ac} +\cc ,
\doneEQ
and the chiral currents have the form 
\EQ
\curr{C}{}{a}= \sum\nolimits_{\KY{}{},\KV{}{}}
( \prodsKY{2}{bcde}{} \der{b}\sF{}{de} 
+\tfrac{1}{5} \der{b} \prodsKY{2'}{bcde}{} \sF{}{de} )
\cLie{\KV{}{}}\sF{}{ae} 
+\cc .
\doneEQ

\subsection{ Graviton fields }
\label{spintwo}

The tensorial form of a spin $s=2$ field consists of 
a real trace-free tensor with Riemann symmetries, 
\EQ
\C{}{abcd}= \C{}{[cd][ab]}
=\vol{AB}{}\vol{CD}{}\csphi{}{A'B'C'D'} + \cc ,\ 
\C{d}{adc}= \duC{d}{adc}= 0 ,
\doneEQ
representing the graviton field strength, 
where the dual tensor is 
$\duC{}{abcd}=\i\vol{AB}{}\vol{CD}{}\csphi{}{A'B'C'D'} + \cc$. 
The graviton field equations 
\EQ
\coder{a} \C{}{abcd}(x) =\coder{a} \duC{}{abcd}(x) =0 
\doneEQ
are analogous to Maxwell's equations,
but with conformal scaling weight $w=1$. 
Decomposition of $\C{}{abcd}$ gives self-dual and antiself-dual parts 
\EQs
&
\sC{}{abcd} =\tfrac{1}{2}(\C{}{abcd}-\i\duC{}{abcd}) 
= \vol{AB}{}\vol{CD}{}\csphi{}{A'B'C'D'}, 
\\&
\aC{}{abcd} =\tfrac{1}{2}(\C{}{abcd}+\i\duC{}{abcd}) =\overline{\sC{}{abcd}}
= \vol{A'B'}{}\vol{C'D'}{}\sphi{}{ABCD} .
\doneEQs

The scaling and duality rotation symmetries are given by 
$\symm{S}{}{abcd} =\C{}{abcd}$, 
$\symm{}{}{abcd} =\duC{}{abcd} = *\symm{S}{}{abcd}$, 
and the spacetime symmetries
depending on real conformal \Kvec/s $\KV{c}{}$ 
are given by 
\EQ
\symm{K}{}{abcd} =
\cwLie{1}{\KV{}{}} \C{}{abcd} ,\quad
\symm{}{}{abcd} =
\cwLie{1}{\KV{}{}} \duC{}{abcd} = *\symm{K}{}{abcd} .
\doneEQ
The chiral symmetries have the lengthy form 
\mEQ
\symm{C}{}{abcd}  = \sum\nolimits_{\KY{}{}} (
\prodsKY{4}{ghjk}{e[b|f[d} 
\der{c]} \der{|a]} \coder{f}\coder{e} \sC{}{ghjk}
+\\
\tfrac{32}{9} 
\der{\indsymm([a|}\prodsKY{4}{ghjk}{e|b]^+f[d} 
\der{c])} \coder{f}\coder{e} \sC{}{ghjk}
+
\tfrac{2}{9} 
\nder{2}{\indsymm([a|}{e} \prodsKY{4'}{ghjk}{e|b]f[d}
\der{c])}\coder{f} \sC{}{ghjk}
\\
+
\tfrac{4}{9} 
\nder{2}{\indsymm([c|[a|}{} \prodsKY{4'}{ghjk}{e|b]^+f|d]^+)}
\coder{e}\coder{f} \sC{}{ghjk} 
+\\
\tfrac{8}{21}
\nder{3}{\indsymm([a|[c|}{e}  \prodsKY{4''}{ghjk}{f|d]^+e|b])}
\coder{f} \sC{}{ghjk}
+
\tfrac{1}{21}
\nder{4}{[a|[c|}{ef} \prodsKY{4'''}{ghjk}{f|d]e|b]} 
\sC{}{ghjk} )
\\
+\cc
\donemEQ
depending on self-dual conformal \KYten/s $\KY{ab}{}$, 
where
\EQs
\prodsKY{4}{ghjkedcb}{} &=
\KY{gh}{}\KY{jk}{}\KY{ed}{}\KY{cb}{} ,
\\
\prodsKY{4'}{ghjkedcb}{} &= 
\KY{gh}{}\KY{jk}{}( \KY{cb}{}\KY{ed}{} -12 \KY{c[e}{}\KY{d]^+b}{} )
\\
\prodsKY{4''}{ghjkedcb}{} &=
( (\KY{cb}{}\KY{ed}{}-4\KY{c[e}{}\KY{d]^+b}{})\KY{gh}{}
-8\KY{c[g}{}\KY{h]^+b}{}\KY{ed}{} )\KY{jk}{}
\\
\prodsKY{4'''}{ghjkedcb}{} &=
\KY{gh}{}\KY{ed}{} \KY{cb}{}\KY{jk}{}
+\tfrac{32}{3} \KY{e[g}{}\KY{h]^+d}{} \KY{c[j}{}\KY{k]^+b}{}
\notag\\
&\quad 
-8( \KY{e[g}{}\KY{h]^+d}{}\KY{jk}{} +\KY{e[j}{}\KY{k]^+d}{}\KY{gh}{} )\KY{cb}{}
\doneEQs
are product tensors arising from the tensorial form of 
the factorization \eqref{KYproduct} 
of rank-four self-dual conformal \KYten/s. 
Here 
\EQs
& 
\nder{2}{ab}{} = 
\der{a}\der{b} -\tfrac{1}{4}\flat{ab} \der{c}\coder{c} ,\quad
\nder{3}{abc}{} = 
\der{a}\der{b}\der{c} -\tfrac{1}{2}\flat{(ab}\der{c)}\der{d}\coder{d} ,
\\
&
\nder{4}{abcd}{} = 
\der{a}\der{b}\der{c}\der{d} 
-\tfrac{3}{4}\flat{(ab}\der{c}\der{d)}\der{e}\coder{e}
+\tfrac{1}{16}\flat{(ab}\flat{cd)}(\der{e}\coder{e})^2 ,
\doneEQs
are the trace-free derivatives, 
and the index operator $\indsymm$ is defined by 
symmetrization over two pairs of skew indices $[ab][cd]$. 

The spacetime currents and zilch currents are given by 
\EQs
\curr{K}{}{a} &= \sum\nolimits_{\KV{}{}}
\KV{c}{}\KV{}{e}\KV{}{f} \sC{}{abcd} \aC{bedf}{} +\cc ,
\\
\curr{Z}{}{a} &= \sum\nolimits_{\KV{}{}}
\i \KV{c}{}\KV{}{e}\KV{}{f} \aC{bedf}{} \cwLie{1}{\KV{}{}}\sC{}{abcd} 
+\cc ,
\doneEQs
and the chiral currents have the form 
\mEQ
\curr{C}{}{a} = 
\sum\nolimits_{\KY{}{},\KV{}{}}
( \prodsKY{4''}{cdefghbj}{} \der{b}\sC{}{cdef} 
+\tfrac{1}{3} \der{b}\prodsKY{4'''}{bjghcdef}{} \sC{}{cdef} )
\cwLie{1}{\KV{}{}}\sC{}{ajgh}
\\
+\cc .
\donemEQ

\subsection{ Neutrino and gravitino fields }
\label{spinhalf}
We first recall the gamma matrices \cite{Penrosebook}
\EQ
\gmatr{}{a} =
\sqrt{2}\begin{pmatrix} 0 & \sodder{aB'}{C} \\ \sodder{aB}{C'} & 0 
\end{pmatrix} 
,\quad
\gmatr{}{5} 
=\tfrac{1}{4!}\vol{}{abcd}\gmatr{}{a}\gmatr{}{b}\gmatr{}{c}\gmatr{}{d}
=\begin{pmatrix} \i & 0 \\ 0 & -\i \end{pmatrix} .
\doneEQ

The Dirac 4-spinor form of a spin $s=\sfrac{1}{2}$ neutrino field 
is represented by a Majorana spinor 
satisfying the massless Dirac field equation
\EQ
\M{}{} =\fourspin{\csphi{}{C'}}{\sphi{}{C}} ,\quad
\gmatr{a}{}\der{a} \M{}{} =0 .
\doneEQ
The scaling and duality rotation symmetries are simply
$\symm{S}{}{} =\M{}{}$, 
$\symm{}{}{} =\gmatr{}{5}\M{}{} = \gmatr{}{5}\symm{S}{}{}$, 
while the spacetime symmetries are given by 
\EQ
\symm{K}{}{} =
\cwLie{-1}{\KV{}{}} \M{}{} ,\quad
\symm{}{}{} =
\cwLie{-1}{\KV{}{}} \gmatr{}{5}\M{}{} = \gmatr{}{5}\symm{K}{}{} 
\doneEQ
which depend on real conformal \Kvec/s $\KV{a}{}$. 
Note $\cwLie{-1}{\KV{}{}}=\sLie{\KV{}{}}$ 
appears due to the conformal scaling weight $w=-1$ of 
the Dirac operator $\gmatr{a}{}\der{a}$. 
The chiral symmetries have the simple form 
\EQ
\symm{C}{}{} = \sum\nolimits_{\KY{}{}}
\rKY{ab}{} \gmatr{}{a}\der{b} \M{}{}
+\tfrac{1}{3}( \der{b}\rKY{ab}{} - \der{b}{*\rKY{ab}{}} \gmatr{}{5} )
\gmatr{}{a} \M{}{} 
\doneEQ
depending on real conformal \KYten/s 
$\rKY{ab}{}= \tfrac{1}{2}( \KY{ab}{}+\cKY{ab}{} )$. 

The spacetime currents reduce here to 
$\curr{}{}{a}= \M{}{}\ad \gmatr{}{a} \M{}{}$, 
which physically describes the neutrino particle density current,
where $\dagger$ denotes the transpose spinor. 
The zilch currents are given by 
\EQ
\curr{Z}{}{a}= 
(\cwLie{-1}{\KV{}{}}\M{}{})\ad \gmatr{}{5}\gmatr{}{a} \M{}{} , 
\doneEQ
while the chiral currents have the form 
\EQ
\curr{C}{}{a}= \sum\nolimits_{\KY{}{},\KV{}{}}
(\cwLie{-1}{\KV{}{}}\M{}{})\ad( 
\rKY{bc}{} \gmatr{}{a}\gmatr{}{c}\der{b} \M{}{}
+\tfrac{1}{3}( \der{b}\rKY{bc}{} -\der{b}{*\rKY{bc}{}} \gmatr{}{5} )
\gmatr{}{a}\gmatr{}{c} \M{}{} ) . 
\doneEQ

Finally, 
a spin $s=\sfrac{3}{2}$ gravitino field is represented by 
a hybrid antisymmetric tensor/Majorana 4-spinor of the form 
\EQ
\M{}{ab} 
=\fourspin{\vol{AB}{}\csphi{}{A'B'C'}}{\vol{A'B'}{}\sphi{}{ABC}} ,\quad
\gmatr{a}{}\M{}{ab}=0 ,
\doneEQ
with left and right handed parts 
$\hM{\pm}{}{ab} =\tfrac{1}{2}(1\mp\i\gmatr{}{5})\M{}{ab} 
= \tfrac{1}{2}(1\mp\i*)\M{}{ab}$,
related by conjugation, 
$\hM{\pm}{}{ab} \equiv \overline{\hM{\mp}{}{ab}}$. 
The gravitino field equation is 
\EQ
\gmatr{c}{}\der{c} \M{}{ab} =0,\quad \text{or equivalently,}\quad
\coder{a}\M{}{ab} =0 ,
\doneEQ
the latter being conformally scaling invariant. 

The gravitino scaling and duality symmetries 
as well as the spacetime symmetries
are analogous to those for neutrino fields, 
$\symm{S}{}{ab} =\M{}{ab}$, 
$\symm{K}{}{ab} = \cLie{\KV{}{}} \M{}{ab}$,
while the spacetime and zilch currents are given by 
\EQ
\curr{K}{}{a} = \sum\nolimits_{\KV{}{}}
\KV{}{b}\KV{c}{}(\M{bd}{})\ad \gmatr{}{c}\M{}{ad} ,\quad
\curr{Z}{}{a} = \sum\nolimits_{\KV{}{}}
\KV{}{b}\KV{c}{}(\M{bd}{})\ad \gmatr{}{5}\gmatr{}{c}\cLie{\KV{}{}}\M{}{ad} .
\doneEQ
In contrast, 
the chiral symmetries and currents have a more complicated form
than those in the neutrino case, 
\mEQ
\symm{C}{}{ab} = \sum\nolimits_{\KY{}{}} ( 
\prodsKY{3}{fged}{c[b} \der{a]}\der{d}\coder{c}\gmatr{}{e}\sM{}{fg} 
+\tfrac{6}{7} \der{c}\prodsKY{3}{fgedc}{[b} 
\der{a]}\der{d}\gmatr{}{e}\sM{}{fg} 
+\\
\tfrac{12}{7}\der{[a}\prodsKY{3}{fgedc}{b]^+} 
\der{c}\der{d}\gmatr{}{e}\sM{}{fg} 
+\tfrac{2}{7}
\nder{2}{cd}{}\prodsKY{3'}{fgedc}{[b} \der{a]^+}\gmatr{}{e}\sM{}{fg} 
+\\
\tfrac{1}{7}
\nder{2}{c[a}{}\prodsKY{3'}{fgedc}{b]} \der{d}\gmatr{}{e}\sM{}{fg} 
+\tfrac{12}{35} \nder{3}{cd[a}{}\prodsKY{3'''}{fgedc}{b]} 
\gmatr{}{e}\sM{}{fg} 
) +\cc ,
\donemEQ
\mEQ
\curr{C}{}{a} = \sum\nolimits_{\KY{}{},\KV{}{}} (
\prodsKY{3''}{fgdebc}{}(\der{b}\sM{}{fg})\ad 
+\tfrac{4}{7} \der{b}\prodsKY{3'}{fgbcde}{} (\sM{}{fg})\ad )
\gmatr{}{c}\gmatr{}{a} \cLie{\KV{}{}}\sM{}{de} 
\\
+\cc ,
\donemEQ
owing to the presence of the product tensors 
\EQs
& 
\prodsKY{3}{fgedcb}{} 
=\KY{fg}{}\KY{ed}{}\KY{cb}{} ,
\\
&
\prodsKY{3'}{fgedcb}{} 
=(\KY{fg}{}\KY{ed}{}-8\KY{f[e}{}\KY{d]^+g}{}) \KY{cb}{} ,
\\
&
\prodsKY{3''}{fgedcb}{} 
=(\KY{fg}{}\KY{ed}{}-4\KY{f[e}{}\KY{d]^+g}{}) \KY{cb}{} 
-4\KY{c[e}{}\KY{d]^+b}{}) \KY{fg}{} ,
\\
&
\prodsKY{3'''}{fgedcb}{} 
=\KY{fg}{} (\KY{ed}{}\KY{cb}{}-\tfrac{4}{3}\KY{c[e}{}\KY{d]^+b}{})  ,
\doneEQs
which are associated with the factorization \eqref{KYproduct} 
of rank-three self-dual conformal \KYten/s.

\section{ Concluding remarks }

Our results on local currents provide a complete set of 
conserved quantities on Minkowski space for the propagation of
electromagnetic and graviton fields 
described using tensorial field strengths,
as well as massless neutrino and gravitino fields 
described in Dirac 4-spinor form. 
In addition, our results on local symmetries hold interest 
for the study of connections between 
symmetry operators and separation of variables 
for these physical field equations. 
Local symmetries and currents, moreover, 
are important in the investigation of nonlinear interactions allowed 
for massless fields \cite{Anco,Henneaux}.

A classification of further symmetries and currents 
involving the familiar electromagnetic and graviton potentials
will be given elsewhere 
by an application of cohomology results for the solution jet space of
the massless spin $s$ field equation.

\end{document}